\documentclass[fleqn,twoside]{article}
\usepackage{espcrc2}
\usepackage{graphicx}
\usepackage[figuresright]{rotating}

\newcommand{\AmS}{{\protect\the\textfont2
  A\kern-.1667em\lower.5ex\hbox{M}\kern-.125emS}}
\def\lesssim{\hbox{ \raise.35ex\rlap{$<$}\lower.6ex\hbox{$\sim$}\ }}
\def\lsim{\hbox{ \raise.35ex\rlap{$<$}\lower.6ex\hbox{$\sim$}\ }}
\def\gtrsim{\hbox{ \raise.35ex\rlap{$>$}\lower.6ex\hbox{$\sim$}\ }}
\def\xrightarrow#1#2#3#4{\,\lower#1pt\hbox{$\stackrel{\stackrel
{\displaystyle #2}%
{\hbox to #3cm{\rightarrowfill}}}{#4}$}\,}
\title{The interplay between high energy physics and cosmology: an example}
\author{M. Sakellariadou\address{Division of Astrophysics, Astronomy, 
and Mechanics, Department of
Physics,\\ University of Athens, Panepistimiopolis, GR-15784 Zografos, 
Greece}
\thanks{msakel@cc.uoa.gr}
}
\begin{document}
\begin{abstract}
Cosmology and high energy physics are two closely connected
areas. In this lecture I present an example of their
rich interplay. 
\vspace{1pc}
\end{abstract}

\maketitle
\section{Introduction}

The cosmology of the early universe is a fast progressing area of
physics.  This remarkable progress has been mainly achieved by
employing high energy physics models.

Most aspects of high energy physics beyond the standard model can only
be tested by going to very high energies, which are by far greater
than those accessible by present, or even future, terrestrial
accelerators.  The rich interplay between particle physics and
cosmology has offered a promising approach to {\sl experimentally}
test new theories of fundamental forces.

Thus, high energy physics models give us the means to formulate
scenarios of the evolution of the early universe, while by confronting
the predictions of these scenarios against cosmological data, one can
constrain the parameters, or even falsify theories of fundamental
forces.

An example of the interplay between cosmology and high energy 
physics is the aim of my talk.

\section{The plot: Topological Defects and CMB data}

\subsection{Topological Defects}

Many particle physics models of matter admit solutions which
correspond to a class of topological defects.  
Under the hypothesis that we understand properly, both unification of 
forces, as well as big bang cosmology, we  expect that
topological defects could have formed naturally during phase
transitions followed by spontaneously broken symmetries, in the early
stages of the evolution of the universe.  Among the various types
of topological defects, some lead to disastrous consequences for 
cosmology and thus, they are undesired, while some others may play a 
useful r\^ole.

Spontaneous symmetry breaking is an old idea, described within the
particle physics context in terms of the Higgs field.
The Symmetry is called Spontaneously Broken (SSB) if the ground state
is not invariant under the full symmetry of the Lagrangian
density. Thus, the vacuum expectation value of the Higgs field is
nonzero. In quantum field theories, broken symmetries are restored at
high enough temperatures.

In three spatial dimensions, four different kinds of topological
defects can arise. The criterion for their formation during a SSB
phase transition, as well as the determination of their type, both
depend on the topology of the vacuum manifold ${\cal M}$. The
properties of ${\cal M}$ are usually described by the $n^{\rm th}$
homotopy group $\pi_n({\cal M})$.  If ${\cal M}$ has disconnected
components, or equivalently if $\pi_0({\cal M})\neq I$, then 
two-dimensional defects, called {\sl domain walls}, form.  The
spacetime dimension $d$ of the defects is given in terms of the order
of the nontrivial homotopy group by $d=4-1-n$. If ${\cal M}$ is not
simply connected, in other words if ${\cal M}$ contains loops which
cannot be continuously shrunk into a point, then {\sl cosmic strings}
form. A necessary, but not sufficient, condition for the existence of
stable strings is that the fundamental group $\pi_1({\cal M})$ of
${\cal M}$, is nontrivial, or ${\cal M}$ is multiply connected. Cosmic strings
are line-like defects, $d=2$. If ${\cal M}$ contains unshrinkable
surfaces, then {\sl monopoles} form.  Finally, if ${\cal M}$ contains
noncontractible three-spheres then event-like defects, {\sl textures},
form for which $n=3, ~d=0$.

Depending on whether the symmetry is local (gauged) or global
(rigid), topological defects are called local or global. The energy of
local defects is strongly confined, while the gradient energy of
global defects is spread out over the causal horizon at defect
formation.

\subsection{Cosmic Microwave Background Temperature Anisotropies}

The Cosmic Microwave Background (CMB) temperature anisotropies provide
a powerful test for theoretical models aiming at describing the early
universe.  The characteristics of the CMB anisotropy multipole
moments, and more precisely the position and amplitude of the acoustic
peaks, as well as the statistical properties of the CMB anisotropies,
can be used to discriminate among theoretical models, as well as to
constraint the parameters space. CMB anisotropies are characterized by
their angular power spectrum $C_\ell$, which is the average value of
the square of the coefficients of a spherical harmonic decomposition
of the measured CMB pattern.

The predictions of the defects models regarding the characteristics of
the CMB spectrum are:\\ $\bullet$ Global ${\cal O}(4)$ textures
predict the position of the first acoustic peak at $\ell\simeq 350$
with an amplitude $\sim 1.5$ times higher than the Sachs-Wolfe
plateau~\cite{rm}.\\ $\bullet$ Global ${\cal O}(N)$ textures in the
large $N$ limit lead to a quite flat spectrum, with a slow decay after
$\ell \sim 100$~\cite{dkm}. Similar are the predictions of other
global ${\cal O}(N)$ defects~\cite{clstrings,num}.\\ $\bullet$ Local
cosmic strings predictions are not very well established and range
from an almost flat spectrum~\cite{acdkss} to a single wide bump at
$\ell \sim 500$~\cite{mark} with extremely rapidly decaying tail.

The position and amplitude of the acoustic peaks, as found by the CMB
measurements~\cite{maxi,boom,dasi,wmap}, are in disagreement with the
predictions of topological defects models.

In addition, topological defects predict nongaussian statistics of
the CMB anisotropies. One could address the question whether the
inflaton field can also give some nongaussian signatures.  It is
often assumed that the initial state of the perturbations of the
inflaton field is the vacuum. In the absence of a theoretical justification
for this assumption, one may relax it. The simplest way to generalise
the vacuum initial state, which contains no privileged scale, is to
consider~\cite{MRS} an initial state with a built-in characteristic
scale. In a band localized around the preferred scale, the state
contains a number of quanta, whereas it is still the vacuum elsewhere.
A robust prediction of such a model is the nongaussian character of
the induced perturbations.  For models with a preferred scale, the
three point (and any higher-order odd-point) correlation function
vanishes, whereas the four-point (and any higher-order even-point)
correlation function does not satisfy Gaussian
statistics~\cite{MRS}. Studying such a model in the context of
single-field inflation, we have shown~\cite{MRS,GMS} that the
nongaussian signature is much smaller than the cosmic variance, thus
undetectable. We have  thus concluded that Gaussian statistics is a
robust prediction of single-field inflation.  From the experimental
point of view, nongaussianity is strongly constrained from the WMAP
measurements~\cite{wmap-nongaus}.

In conclusion, CMB measurements rule out pure topological defects
models as the origin of initial density perturbations; inflation wins
over topological defects.  This leads to a crucial set of questions
concerning high energy physics.  Namely, are topological
defects, and more precisely cosmic strings, allowed at all? We are
basically interested in cosmic strings, since we consider gauge
theories (domain walls and monopoles are dangerous, while textures are
uninteresting~\cite{textures}). How generic is cosmic strings
formation?  Which are the consequences for fundamental theories?  In
what follows, we address these questions.

It is conceivable to consider a mixed perturbation model, in which the
primordial fluctuations are induced by an inflaton field with a
non-negligible cosmic strings contribution.  We have
considered~\cite{bprs} a model in which a network of cosmic strings
evolved independently of any pre-existing fluctuation background,
generated by a standard cold dark matter with a nonzero cosmological
constant inflationary phase. Restricting our attention to the angular
spectrum, we can remain in the linear regime.  Thus,
\begin{equation}
C_\ell =   \alpha     C^{\scriptscriptstyle{\rm I}}_\ell
         + (1-\alpha) C^{\scriptscriptstyle{\rm S}}_\ell~,
\label{cl}
\end{equation}
where $C^{\scriptscriptstyle{\rm I}}_\ell$ and $C^{\scriptscriptstyle
{\rm S}}_\ell$ denote the (COBE normalized) Legendre coefficients due
to adiabatic inflation fluctuations and those stemming from the string
network respectively. The coefficient $\alpha$ in Eq.~(\ref{cl}) is a
free parameter giving the relative amplitude for the two
contributions.  One has to compare the $C_\ell$, given by
Eq.~(\ref{cl}), with data obtained from CMB measurements.  Already
MAXIMA~\cite{maxi}, BOOMERanG~\cite{boom} and DASI~\cite{dasi}
experiments imposed~\cite{bprs} an upper limit on the cosmic
strings contribution to the CMB, which is $\lesssim 18\%$.  Clearly,
the limit set by the Wilkinson Microwave Anisotropy Probe (WMAP)
measurements~\cite{wmap} should be stronger. A recent Bayesian
analysis in a three dimensional parameter space~\cite{pogosian} has
shown that a cosmic strings contribution to the primordial
fluctuations higher than $9\%$ is excluded up to $99\%$ confidence
level.

\section{The plot thickens: Genericity of cosmic strings formation in
SUSY GUTs}

The natural question one has to address is how generic cosmic strings
formation is. Clearly the answer to this question depends on the
framework we are placed in. This issue has been studied in detail
within Supersymmetric Grand Unified Theories (SUSY GUTs) in
Ref.~\cite{rjs}.

Grand Unified Theories  imply that our universe has undergone a
series of phase transitions associated with the SSB of the GUT gauge
group ${\rm G}_{\rm GUT}$ down to the standard model gauge group ${\rm
G}_{\rm SM} = {\rm SU}(3)_{\rm C} \times {\rm SU}(2)_{\rm L} \times
{\rm U}(1)_{\rm Y}$ at $M_{\rm GUT} \sim 3 \times 10^{16}$ GeV.
There might be one, more than one, or no intermediate
symmetry group between ${\rm G}_{\rm GUT}$ and ${\rm G}_{\rm SM}$. As
a cosmological consequence of these SSB patterns one obtains the
formation of topological defects via the Kibble
mechanism~\cite{kibble}.

Spontaneous symmetry breaking schemes which lead to the formation of
monopoles or domain walls are ruled out since they are incompatible
with our universe, unless an inflationary era took place after their
formation. We cannot constrain SSB schemes with texture formation,
since this class of defects cannot play a significant r\^ole in
cosmology~\cite{textures}.

The particle physics Standard Model (SM) has been tested to a very
high precision, however experimental data, and in particular evidence
of neutrino masses~\cite{SK,SNO,kamland}, show that one should go
beyond this model. An extension of the SM gauge group is realised in
the framework of Supersymmetry (SUSY), which is at present the only
viable theory for solving the gauge hierarchy problem. In addition,
within SUSY GUTs the gauge coupling constants of the strong, weak and
electromagnetic interactions meet at a single point $M_{\rm GUT}
\simeq (2-3) \times 10^{16}$ GeV.  Finally, SUSY GUTs can provide the
scalar field to play the r\^ole of an inflaton field, explain the
baryon asymmetry of the universe, and provide a
candidate (the lightest superparticle) for cold dark matter. 

In Ref.~\cite{rjs}, we have considered all possible SSB schemes from a
large gauge group ${\rm G_{GUT}}$ down to ${\rm G_{SM}}\times Z_2$, in
the context of SUSY GUTs. $Z_2$ is a sub-group of the ${\rm U}(1)_{\rm
B-L}$ gauge symmetry and it plays the r\^ole of R-parity. The
requirement of an unbroken R-parity down to low energies guarantees
proton stability.  We have limited the choice of ${\rm G_{GUT}}$
to simple gauge groups which contain ${\rm G_{SM}}$, have a
complex representation, are anomaly free, and whose rank is not higher
than 8; the main conclusions remain qualitatively unaffected for
groups of higher rank. We have studied the homotopy group of the
vacuum manifold to find the type of defects formed, if any.

We have considered an exhaustive list of possible embeddings of ${\rm
G_{SM}}$ in ${\rm G_{GUT}}$ and we have examined whether defects are
formed during the SSB patterns and of which kind they are. To get rid
of the undesired defects, basically monopoles, we have employed an era
of standard hybrid inflation after their formation.
Moreover, we have considered a mechanism of baryogenesis via
leptogenesis, which can be thermal or nonthermal one.  In the case of
nonthermal leptogenesis, U(1)$_{\rm (B-L)}$ is a sub-group of the GUT
gauge group, G$_{\rm GUT}$, and B-L is broken at the end or after
inflation. In the case of thermal leptogenesis, B-L is broken
independently of inflation. If leptogenesis is thermal and B-L is
broken before the inflationary era, then one should check whether the
temperature at which B-L is broken, which will define the mass of the
right-handed neutrinos, is smaller than the reheating temperature
which should be lower than the limit imposed by the gravitino.

We have then asked how generic is cosmic strings formation after
hybrid inflation, within these schemes.  Only if we relax the
requirement that the gauged ${\rm B-L}$ symmetry is broken at the end
of inflation, there are a few ($\sim 2\%$) SSB schemes without cosmic
strings formation, otherwise cosmic strings formation is generic.  The
number of SSB schemes with no cosmic strings formation increases
($\sim 15\%$), if we accept SSB schemes with broken R-parity, but
then it remains an open question how to stabilise the proton.

We have found~\cite{rjs} that cosmic strings formation is sometimes
accompanied by the formation of embedded strings which however are
topologically, and in general also dynamically, unstable~\cite{periv}.

\section{The plot unfolds: Supersymmetric Hybrid Inflation}

One of the main questions within our list is indeed answered.
Cosmic strings are generically formed in the framework of SUSY GUTs.
However, we stated earlier that strong constraints are placed in their
contribution to the CMB power spectrum. Thus, the obvious question to
address at this point, is whether we can constrain the parameter's
space of the models with strings at the end of the last inflationary
era, so that their contribution to the CMB is within the allowed
window. Answering to this question will allow us, at least partially,
to find the class of natural inflationary model, if any. These issues
have been addressed in Refs.~\cite{jm1,jm2,jm3}.

The inflationary paradigm offers the most appealing approach for
describing the early stages of the evolution of our universe.
Inflation essentially consists of a phase of accelerated expansion
which took place at a very high energy scale. Inflation requires the
existence of a slowly rolling scalar field, while inflation will cease
whenever slow-roll conditions are violated.  Inflation comes to
complete the standard Big Bang model and offers an explanation for the
initial density fluctuations leading to the observed structure
formation and the measured anisotropies of the CMB. However, inflation
is faced with two questions, namely how generic is the onset of
inflation and which is a natural model of inflation. We have 
found~\cite{calzsakel1} that the onset of inflation requires some
special initial conditions, which however may  be the likely outcome
of quantum events occurred before the inflationary
era~\cite{calzsakel2}. To find a natural model of inflation,
consistent with high energy physics models and cosmological data
seems to be less trivial.

To describe the early evolution of our universe, at energies below
Planck scale, one should consider an effective N=1 Supergravity
(SUGRA).  This implies that inflationary models should be constructed
in the framework of SUGRA, since the inflationary scale is
$V^{1/4}\lesssim 4\times 10^{16}$ GeV. However, is is difficult to
implement slow-roll inflation within SUGRA. More precisely, the
positive false vacuum of the inflaton field breaks spontaneously
global supersymmetry; it gets restored after the end of the
inflationary era, when the field rolls to the true vacuum. In SUGRA
the SUSY breaking is transmitted to all fields by gravity, thus
any scalar field gets a soft mass given by
\begin{equation}
m_{\rm soft}^2\sim 8\pi V/M_{\rm Pl}^2\sim H^2~,
\end{equation}
where $H$ is the expansion rate during inflation, and $M_{\rm Pl}$
denotes the reduced Planck mass. One has to use fine-tuning 
to avoid such a large soft mass for the scalar field which plays the
r\^ole of the inflaton.
This is known as the problem of ``Hubble-induced mass''. 

In a supersymmetric theory, the tree-level potential is the sum of an
F-term and a D-term, which have different properties. In all proposed
inflationary models one of these two terms is the dominant
one.  It was shown~\cite{dterm}, that the ``Hubble-induced mass''
problem comes from F-term interactions and it may be avoided if
we consider the vacuum energy as being dominated by nonzero D-terms of
some superfields. Inflationary models where the potential is dominated
by nonvanishing D-terms emerge naturally in theories with either an
anomalous or an nonanomalous gauge U(1) symmetry which incorporates a
Fayet-Iliopoulos term. In D-term inflation the masses of the scalar
fields depend on their gauge charges. More precisely, in D-term
inflation, the inflaton field is a singlet under gauge symmetry, thus
the curvature of the inflaton potential is small. Since in addition,
D-term inflation can easily be implemented in string theory, this
class of models gained a lot of interest.

\subsection{F-term Inflation}

F-term inflation can be accommodated in a SSB scheme, where a GUT gauge 
group is broken down to the standard model gauge group at an 
energy scale $M_{\rm GUT}$ according to
\begin{equation}
G_{\rm GUT} \stackrel{M_{\rm GUT}}{\hbox to 0.8cm {\rightarrowfill}} H_1
\xrightarrow{9}{M_{\rm infl}}{1}{\Phi_+\Phi_-} H_2 {\longrightarrow} 
G_{\rm SM}~,
\end{equation}
where $\Phi_+, \Phi_-$ is a pair of GUT Higgs superfields in
nontrivial complex conjugate representations, which lower the rank of
the group by one unit when acquiring nonzero VEV. The inflationary phase 
takes place at the beginning of the SSB  
$H_1\stackrel{M_{\rm infl}}{\longrightarrow} H_2$. F-term inflation 
can be chosen as the hybrid inflationary model introduced in the SSB schemes
studied in the previous section. Thus, generically, one expects the
formation of cosmic strings (accompanied sometimes by embedded strings),
at the end of the inflationary era.

F-term inflation is based on the supersymmetric renormalisable 
superpotential
\begin{equation}\label{superpot}
W_{\rm infl}^{\rm F}=\kappa S(\Phi_+\Phi_- - M^2)~,
\end{equation}
where $S, \Phi_+, \Phi_-$ are chiral superfields, and $\kappa$,
$M$ are two constants.  The scalar potential reads
\begin{eqnarray}
\label{scalpot1}
V(\phi_+,\phi_-,S)&=&
|F_{\Phi_+}|^2+|F_{\Phi_-}|^2+|F_S|^2 \nonumber\\
&&+{1\over 2}\sum_a g_a^2 D_a^2~.
\end{eqnarray}
The F-term is such that $F_{\Phi_i} \equiv |\partial W/\partial
\Phi_i|_{\theta=0}$, where we take the scalar component of the
superfields once we differentiate with respect to $\Phi_i=\Phi_+,
\Phi_-, S$. The D-term is
\begin{equation}
D_a=\bar{\phi}_i\,{(T_a)^i}_j\,\phi^j +\xi_a~,
\end{equation}
with $a$ the label of the gauge group generators $T_a$, $g_a$ the
gauge coupling, and $\xi_a$ the Fayet-Iliopoulos term. By
definition, in the F-term inflation the real constant $\xi_a$ is zero;
it can only be nonzero if $T_a$ generates a U(1) group.

In the context of F-term hybrid inflation, the F-terms give rise to
the inflationary potential energy density, while the D-terms are flat
along the inflationary trajectory, thus one may neglect them during
inflation.

The potential has one valley of local minima, $V=\kappa^2 M^4$, for
$S> M $ with $\phi_+ = \phi_-=0$, and one global supersymmetric
minimum, $V=0$, at $S=0$ and $\phi_+ = \phi_- = M$. Imposing initially
$ S \gg M$, the fields quickly settle down the valley of local minima.
Since in the slow roll inflationary valley, the ground state of the
scalar potential is nonzero, SUSY is broken.  In the tree level,
along the inflationary valley the potential being constant, it is
perfectly flat. A slope along the potential can be generated by
including the one-loop radiative corrections. Thus, the scalar
potential gets a little tilt which helps the inflaton field $S$ to
slowly roll down the valley of minima. The one-loop radiative
corrections to the scalar potential along the inflationary valley,
lead to an effective
potential~\cite{DvaShaScha,Lazarides,SenoSha,jm1,jm2,jm3}
\begin{eqnarray}
\label{VexactF}
V_{\rm eff}^{\rm F}(|S|)&=&\kappa^2M^4\big\{1+\frac{\kappa^2
\cal{N}}{32\pi^2}\big[2\ln\frac{|S|^2\kappa^2}{\Lambda^2}\nonumber\\
&&+\big({|S|^2\over M^2}+1\big)^2\ln\big(1+{M^2\over |S|^2})\big)\nonumber\\
&&+\big({|S|^2\over M^2}-1\big)^2\ln\big(1-{M^2\over |S|^2}\big)\big]\big\}
~;
\end{eqnarray}
$\Lambda$ is a renormalisation scale 
and $\cal{N}$ stands for the dimensionality
of the representation to which the complex scalar components $\phi_+,
\phi_-$ of the chiral superfields $\Phi_+, \Phi_-$ belong. 

Considering only large angular scales, one can study the contributions
to the CMB temperature anisotropies by analytical methods. In
Refs.~\cite{jm1,jm2}, we have calculated explicitly the Sachs-Wolfe
effect. The quadrupole anisotropy has one
contribution coming from the inflaton field, calculated using
Eq.~(\ref{VexactF}), and one contribution coming from the cosmic
strings network, given by numerical simulations~\cite{ls}. Fixing the
number of e-foldings to 60, then for a given gauge group, the inflaton
and cosmic strings contribution to the CMB depend on the
superpotential coupling $\kappa$, or equivalently on the symmetry
breaking scale $M$ associated with the inflaton mass scale, which
coincides with the string mass scale. The total quadrupole anisotropy
has to be normalised to the COBE data.  In Refs.~\cite{jm1,jm2} we have
found that the cosmic strings contribution is consistent with the
CMB measurements, provided
\begin{equation}
M\lsim 2\times 10^{15} {\rm GeV} ~~\Leftrightarrow ~~\kappa \lsim
7\times10^{-7}~.
\end{equation}
Strictly speaking the above condition was found in the context of SO(10) 
gauge group, but the conditions imposed in the context of other gauge
groups are of the same order of magnitude since $M$ is a slowly varying
function of the dimensionality ${\cal N}$ of the representations to which 
the scalar components of the chiral Higgs superfields belong.

The superpotential coupling $\kappa$ is also subject to the gravitino 
constraint which imposes an upper limit to the reheating temperature, 
to avoid gravitino overproduction. Within the framework of SUSY GUTs and
assuming a see-saw mechanism to give rise to massive neutrinos, the 
inflaton field will decay during reheating into pairs of right-handed 
neutrinos.  This constraint on the reheating temperature can be converted
to a constraint on the parameter $\kappa$. The gravitino constraint on
$\kappa$ reads~\cite{jm1,jm2} $\kappa \lesssim 8\times
10^{-3}~$, which is a weaker constraint.

Concluding, F-term inflation leads generically to cosmic strings
formation at the end of the inflationary era. The cosmic strings
formed are of the GUT scale. This class of models can be compatible
with CMB measurements, provided the superpotential coupling is smaller
than $10^{-6}$.  This tuning on the free parameter $\kappa$ can be
softened if one allows for the curvaton mechanism.  According to the
curvaton mechanism~\cite{lw2002,mt2001}, another scalar field, called
the curvaton, could generate the initial density perturbations whereas
the inflaton field is only responsible for the dynamics of the
universe. The curvaton is a scalar field, that is subdominant during
the inflationary era as well as at the beginning of the radiation
dominated era which follows the inflationary phase. There is no
correlation between the primordial fluctuations of the inflaton and
curvaton fields. Clearly, within supersymmetric theories such scalar
fields are expected to exist. In this case, the coupling $\kappa$ is
only constrained by the gravitino limit. More precisely, assuming the
existence of a curvaton field, there is an additional contribution to
the temperature anisotropies. The WMAP CMB measurements
impose~\cite{jm1,jm2} the following limit on the initial value of the
curvaton field
\begin{equation}
{\cal\psi}_{\rm init} \lsim 5\times 10^{13}\,\left( 
\frac{\kappa}{10^{-2}}\right){\rm GeV}~,
\end{equation}
provided the parameter $\kappa$ is in the range $[10^{-6},~1]$.

\subsection{D-term Inflation}

D-term inflation is derived in SUSY from the superpotential
\begin{equation}
\label{superpotD}
W^{\rm D}_{\rm infl}=\lambda S \Phi_+\Phi_-~,
\end{equation}
where $S, \Phi_-, \Phi_+$ are chiral superfields and $\lambda$ is the
superpotential coupling. 
In D-term inflation, as opposed to F-term inflation, the 
inflaton mass acquires values of the order of Planck mass, and
therefore, the correct analysis must be done in the framework of
SUGRA.

For minimal SUGRA, the effective scalar potential reads~\cite{jm1,jm2,jm3}
$$\kern -2.8cm V^{\rm D-SUGRA}_{\rm eff} 
=\frac{g^2\xi^2}{2}\big\{1+\frac{g^2}{16 \pi^2}$$
\begin{eqnarray}
\label{vDsugra}
&&\kern -.6cm \times\big[2\ln\frac{|S|^2\lambda^2}{\Lambda^2}e^{|S|^2\over
M^2_{\rm Pl}}\nonumber\\ &&\kern -.6cm +\big( {\lambda^2 |S|^2\over
g^2\xi}e^{|S|^2\over M_{\rm Pl}^2} +1\big)^2
\ln\big(1+{ g^2\xi\over\lambda^2 |S|^2 }e^{-{|S|^2\over M_{\rm Pl}^2}}
\big)\nonumber\\ 
&&\kern -.6cm +\big( {\lambda^2 |S|^2\over
g^2\xi}e^{|S|^2\over M_{\rm Pl}^2} -1\big)^2
\ln\big(1-{ g^2\xi\over\lambda^2 |S|^2 }e^{-{|S|^2\over M_{\rm Pl}^2}}
\big)\big]\big\}
\end{eqnarray}

D-term inflation requires a SSB pattern,
\begin{equation}
G_{\rm GUT}\times U(1) \stackrel{M_{\rm GUT}}
{\hbox to 0.8cm{\rightarrowfill}} H \times U(1)
\xrightarrow{9}{M_{\rm infl}}{1}{\Phi_+\Phi_-}
H \rightarrow G_{\rm SM}~.
\end{equation}
Clearly, the symmetry breaking at the end of the inflationary phase
implies that cosmic strings are always formed at the end of D-term
hybrid inflation. This statement, thought to cause a problem for D-term
inflation, since it was claimed~\cite{rj} that in this class of
models, the cosmic strings contribution to the CMB measurements is
constant and dominant. In the literature, one can find a number of 
approaches to avoid cosmic strings formation in the context of D-term 
inflation. For example, one can add a nonrenorlisable term 
in the potential~\cite{shifted}, or add an additional discrete 
symmetry~\cite{smooth}, or consider GUT models based on nonsimple 
groups~\cite{mcg}, or introduce a new pair of charged
superfields~\cite{jaa} so that cosmic strings formation is avoided 
within D-term inflation.

In Refs.~\cite{jm1,jm2,jm3}, we have properly addressed the question
of cosmic strings contribution to the CMB data and we have found that
standard D-term inflation can be compatible with measurements; the
cosmic strings contribution to the CMB is actually
model-dependent. Our most important finding was that cosmic strings
contribution is not constant, nor is it always dominant.
   
More precisely, we have found~\cite{jm1,jm2,jm3} that $g\gtrsim
 2\times 10^{-2}$ is incompatible with the allowed cosmic strings
 contribution to the WMAP measurements.  For $g\lsim 2\times 10^{-2}$,
 the constraint on the superpotential coupling $\lambda$ reads
 $\lambda \lsim 3\times 10^{-5}$.  SUGRA corrections impose in
 addition a lower limit to $\lambda$.  The constraints induced on the
 couplings by the CMB measurements can be expressed as a single
 constraint on the Fayet-Iliopoulos term $\xi$, namely $\sqrt\xi \lsim
 2\times 10^{15}~{\rm GeV}$.

Concluding, standard D-term inflation always leads to cosmic strings
formation at the end of the inflationary era. The cosmic strings
formed are of the GUT scale. This class of models is still compatible
with CMB measurements, provided the couplings are small enough.  As in
the previous class of models, the fine tuning on the couplings can be
softened provided one considers the curvaton mechanism. In this case,
the imposed CMB constraint on the initial value of the curvaton field
reads~\cite{jm3}
\begin{equation}
\psi_{\rm init}\lsim 3\times 10^{14}\left({g\over 10^{-2}}\right) ~{\rm GeV},
\end{equation}
for $\lambda\in [10^{-1}, 10^{-4}]$.

\section{Conclusions}

High energy physics models are used to formulate scenarios of
the evolution of the early universe. Comparing the predictions of these
cosmological scenarios against observational and experimental data we
can in return test the original high energy physics models.

Cosmic strings are generically formed in almost all SSB schemes from a
large gauge group down to the standard model. To be consistent with
CMB measurements, we can place limits on the free parameters (masses
and couplings). Thus, F- as well as D-term inflationary models are
found to be consistent with the data, provided the couplings are small
enough, unless we want to employ the curvaton mechanism.

\section*{Acknowledgements}
It is a pleasure to thank the organisers of the Athens-2004 Workshop
for inviting me to give this talk. I also thank my collaborators with
whom I have worked on the issues addressed here.

\end{document}